\documentclass[apjl]{emulateapj}
\usepackage{amsmath,graphicx,epsfig,multirow,natbib,rotating}
\shorttitle{A Strong Lens Galaxy in a Cluster at \zcluster}
\shortauthors{WONG et al.}

\begin{document}

\newcommand{\hubble}{{\it HST}}
\newcommand{\hubblefull}{{\it Hubble Space Telescope}}
\newcommand{\spitzer}{{\it Spitzer}}
\newcommand{\lya}{Ly$\alpha$}
\newcommand{\kms}{km~s$^{-1}$}
\newcommand{\OII}{\hbox{[{\rm O}\kern 0.1em{\sc ii}]} $\lambda3727$}
\newcommand{\OIII}{\hbox{[{\rm O}\kern 0.1em{\sc iii}]} $\lambda\lambda4959,5007$}
\newcommand{\iab}{$i_{\rm AB}$}
\newcommand{\bcg}{UDS40171}
\newcommand{\zcluster}{$z=1.62$}
\newcommand{\zbcg}{$z_{\rm L}=1.64$}
\newcommand{\zbcgfull}{$z_{\rm L}=1.6406^{+0.0018}_{-0.0050}$}
\newcommand{\zsource}{$z_{\rm S}=2.26$}
\newcommand{\zsourcefull}{$z_{\rm S}=2.2623\pm0.0002$}
\newcommand{\zsourcelya}{$z_{\rm S}=2.25384\pm0.00003$}
\newcommand{\cluster}{IRC 0218}
\newcommand{\Msun}{${\rm M}_{\odot}$}

\title{Discovery of a Strong Lensing Galaxy Embedded in a Cluster at $\MakeLowercase{z}=1.62$\footnotemark[*]}  \footnotetext[*]{Based on observations made with the NASA/ESA Hubble Space Telescope, obtained at the Space Telescope Science Institute, which is operated by the Association of Universities for Research in Astronomy, Inc., under NASA contract NAS 5-26555. These observations are associated with program \#12590.}
\author{
Kenneth C. Wong\altaffilmark{1,9},
Kim-Vy H. Tran\altaffilmark{2},
Sherry H. Suyu\altaffilmark{1},
Ivelina G. Momcheva\altaffilmark{3},
Gabriel B. Brammer\altaffilmark{4},
Mark Brodwin\altaffilmark{5},
Anthony H. Gonzalez\altaffilmark{6},
Aleksi Halkola,
Glenn G. Kacprzak\altaffilmark{7,10},
Anton M. Koekemoer\altaffilmark{4},
Casey J. Papovich\altaffilmark{2},
and Gregory H. Rudnick\altaffilmark{8}}
\altaffiltext{1}{Institute of Astronomy and Astrophysics, Academia
Sinica (ASIAA), P.O. Box 23-141, Taipei 10617, Taiwan}
\altaffiltext{2}{George P. and Cynthia W. Mitchell Institute for
Fundamental Physics and Astronomy, Department of Physics \& Astronomy, Texas A\&M
University, College Station, TX 77843, USA}
\altaffiltext{3}{Astronomy Department, Yale University, New Haven, CT 06511, USA}
\altaffiltext{4}{Space Telescope Science Institute, 3700 San Martin Drive, Baltimore, MD 21218, USA}
\altaffiltext{5}{Department of Physics and Astronomy, University of Missouri, 5110 Rockhill Road, Kansas City, MO 64110, USA}
\altaffiltext{6}{Department of Astronomy, University of Florida, Gainesville, FL 32611, USA}
\altaffiltext{7}{Swinburne University of Technology, Victoria 3122, Australia}
\altaffiltext{8}{Department of Physics and Astronomy, The University of Kansas, Malott room 1082, 1251 Wescoe Hall Drive, Lawrence, KS 66045}
\altaffiltext{9}{EACOA Fellow}
\altaffiltext{10}{Australian Research Council Super Science Fellow}

\begin{abstract}
We identify a strong lensing galaxy in the cluster IRC 0218 (also known as XMM-LSS J02182$-$05102) that is spectroscopically confirmed to be at $z=1.62$, making it the highest-redshift strong lens galaxy known.  The lens is one of the two brightest cluster galaxies and lenses a background source galaxy into an arc and a counterimage.  With {\it Hubble Space Telescope} ({\it HST}) grism and Keck/LRIS spectroscopy, we measure the source redshift to be $z_{\rm S}=2.26$.  Using {\it HST} imaging in ACS/F475W, ACS/F814W, WFC3/F125W, and WFC3/F160W, we model the lens mass distribution with an elliptical power-law profile and account for the effects of the cluster halo and nearby galaxies. The Einstein radius is $\theta_{\rm E}=0.38^{+0.02}_{-0.01}\arcsec $ ($3.2_{-0.1}^{+0.2}$ kpc) and the total enclosed mass is M$_{\rm tot} (< \theta_{\rm E})=1.8^{+0.2}_{-0.1}\times10^{11}~{\rm M}_{\odot}$.  We estimate that the cluster environment contributes $\sim10$\% of this total mass.  Assuming a Chabrier IMF, the dark matter fraction within $\theta_{{\rm E}}$ is $f_{\rm DM}^{{\rm Chab}} = 0.3_{-0.3}^{+0.1}$, while a Salpeter IMF is marginally inconsistent with the enclosed mass ($f_{\rm DM}^{{\rm Salp}} = -0.3_{-0.5}^{+0.2}$).  The total magnification of the source is $\mu_{\rm tot}=2.1_{-0.3}^{+0.4}$.  The source has at least one bright compact region offset from the source center.  Emission from Ly$\alpha$ and \hbox{[{\rm O}\kern 0.1em{\sc iii}]} are likely to probe different regions in the source.
\end{abstract}


\section{Introduction} \label{sec:intro}

Gravitational lensing is a powerful tool for studying the mass
structure of galaxies.  Lensing studies of early-type galaxies (ETGs) at $z<1$
have produced interesting results on their properties.  Their total mass
density profile slope $\gamma'$ (where $\rho(r)\propto r^{-\gamma^{\prime}}$)
depends solely on the surface stellar mass density at fixed redshift,
whereas $\gamma^{\prime}$ of individual galaxies does not evolve
significantly \citep{koopmans2006,barnabe2013,sonnenfeld2013}.  ETGs also favor a heavier stellar initial mass function \citep[IMF;][]{auger2010,sonnenfeld2012}.  Nonetheless, the evolution of the mass distribution of ETGs at $z>1$ is not well-constrained.  Identifying strong lensing galaxies at $z>1$ provides leverage on how ETGs assemble and tests the current cosmological paradigm.

However, strong lensing galaxies are increasingly rare at higher
redshifts due to the evolving galaxy mass function, decreasing
background volume, and decreasing lensing efficiency.  Distant lenses
also require high-resolution imaging to separate the source and lensing galaxy, and near-infrared spectroscopy to confirm their redshifts.  \citet{vanderwel2013} reported 
the most distant strong lensing galaxy known thus far at $z_{\rm L}=1.53$ and estimated 
one $z_{\rm L}>1$ system per $\sim200$ arcmin$^{2}$.

Here, we report the discovery of a strong lensing galaxy embedded in a cluster at \zcluster\
\citep{papovich2010,tanaka2010}, making it the highest-redshift strong lens galaxy known \citep[in contrast with the highest-redshift strong lens cluster at $z = 1.75$;][]{gonzalez2012}.  The system is unusual because of the high lens redshift, and the lens being the most massive cluster member. The system lies in the UDS legacy field, which has extensive multi-wavelength observations \citep{lawrence2007,galametz2013,skelton2014}. Spectroscopy taken with the
\hubblefull\ (\hubble) grisms and the {\it Low Resolution Imaging Spectrometer} \citep[LRIS;][]{oke1995} on Keck confirm the source redshift of
\zsource.  We combine the datasets to model the lens, including its
environment.  We assume $\Omega_{m} = 0.3$,
$\Omega_{\Lambda}=0.7$, and $H_{0} = 70$\,km\,s$^{-1}$\,Mpc$^{-1}$.
All quantities in $h_{70}$ units.  At \zcluster, the
angular scale is $1\arcsec = 8.47$\,kpc.  We use AB magnitudes.

\section{Observations} \label{sec:data}

\subsection{Keck/LRIS Spectroscopy} 

Using the Keck/LRIS,  we carried out a spectroscopic
survey of the cluster \cluster on 2012 October 19 \& 20 (NASA/Keck Program ID 48/2012B).  Targets were selected from the
\citet{williams2009} catalog of the Ultra-Deep Survey (UDS) taken as
part of the UKIRT Infrared Deep Sky Survey \citep[UKIDSS;][]{lawrence2007}.  The $K$-selected catalogs reach $5\sigma$-limiting magnitudes in 1.75\arcsec~diameter apertures of $B<27.7$, $R<27.1$, $i<26.8$, $z<25.5$, $J<23.9$, and $K<23.6$.  

We use the 600/4000 grism for the blue side and the 600/10000
grating for the red side of LRIS.  With 1\arcsec~slit widths, the resolutions are 
4.0~\AA~and 4.7~\AA, respectively.  Observing conditions were excellent with $\sim0.6\arcsec$ median seeing.  The total integration time on the lens is 9 hours.  We reduce the spectra following \citet{tran2007}, using
IRAF\footnote{IRAF is distributed by the National Optical Astronomy
Observatories, which are operated by the Association of Universities
for Research in Astronomy, Inc., under cooperative agreement with the
National Science Foundation.} routines with custom software provided
by D. Kelson \citep{kelson2003}.  The wavelength coverage of the
extracted spectra is $3800-5800$~\AA\ (blue side) and
$7000-10000$~\AA\ (red side).  A full analysis of this redshift
survey will be presented in Tran et al. (in preparation).

\begin{figure*}
\plotone{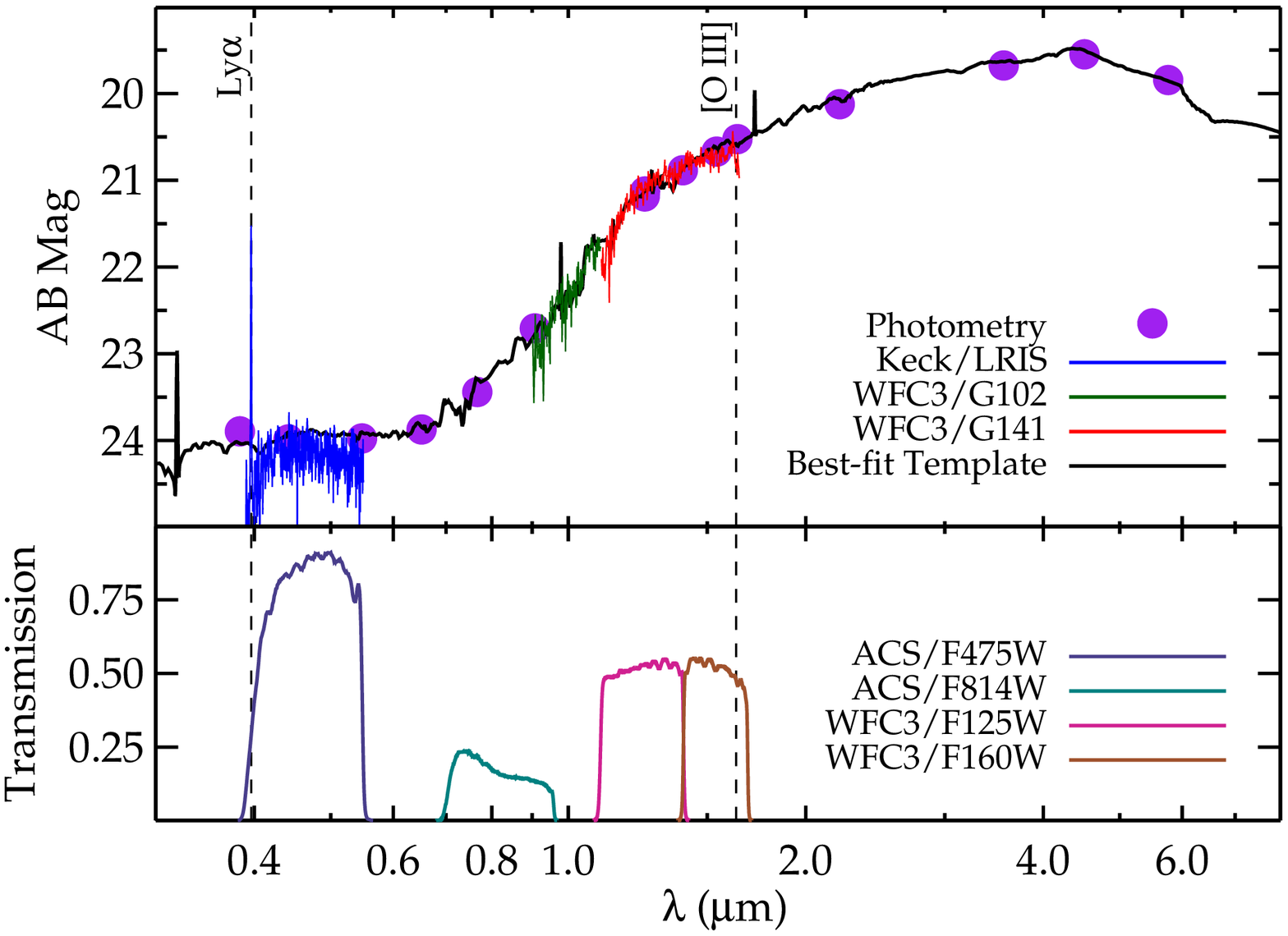}
\caption{{\bf Top:} Total fluxes for the lens system as measured by
3D-HST \citep[object \#31684;][]{skelton2014} that combines
ground-based imaging with \hubble\ and \spitzer\ observations (purple).  The
best-fit SED from EAZY is included (black).  Also shown are the
blended Keck/LRIS spectrum (blue) and the
\hubble/WFC3 G102 (green) and G141 (red) grism spectra for the 
lens.   The spectra are arbitrarily normalized for clarity.  The grism spectra trace the lens galaxy's Balmer break at \zbcg.  The spectra show strong \lya\ and \OIII\ emission from the
source at \zsource.
{\bf Bottom:} Filter transmission
curves for the \hubble\ imaging used in the lens modeling.  A bright
point source corresponding to emission at \zsource\ is visible
in the ACS/F475 (\lya) and WFC3/F160 (\OIII) images.
\label{fig:sed}}
\end{figure*}

We use XCSAO \citep{kurtz1992} with various templates for Lyman-break
and \lya~emitting galaxies from \citet{shapley2003} to
measure a redshift for the blended spectrum composed of the lens
galaxy and the lensed source.  The strong emission detected at
3957~\AA\ (Figure~\ref{fig:sed}, top) has the asymmetric profile
characteristic of \lya\ \citep{shapley2003} and corresponds to a
redshift of \zsourcelya.  The observed \lya\ equivalent width, derived for a continuum fit between $1180-1250$~\AA~rest-frame, is $129.4 \pm 4.6$~\AA.

\subsection{\hubble\ Observations}

The UDS cluster falls in a legacy field that has extensive
multi-wavelength observations, including deep
\hubble\ imaging from CANDELS \cite[$0.06''$/pixel;][]{grogin2011,koekemoer2011}
and G141 spectroscopy from 3D-HST \citep{brammer2012}.
Additional \hubble\ imaging and G102 spectroscopy were obtained in
GO-12590 (PI: C. Papovich).  Figure~\ref{fig:sed} shows the total
fluxes measured for the more massive of the two brightest cluster galaxies (BCG)
from the 3D-HST catalog of the UDS field \citep{skelton2014}, which
includes \spitzer/IRAC.  Hereafter, the BCG refers to this more massive galaxy unless otherwise stated.  The BCG shows excess flux at
$\lambda<5000$~\AA\ due to emission from the lensed source.
At longer wavelengths, the lens galaxy dominates.  Grism spectroscopy
confirms the BCG redshift of \zbcgfull.  We attribute the redshift
difference between the BCG and cluster to peculiar velocities due to
the unrelaxed nature of the cluster \citep[e.g.,][]{martel2014}, consistent with the high fraction of star-forming members \citep{tran2010}.  We use \zcluster~as the lens redshift for cosmological calculations.

\begin{figure}
\plotone{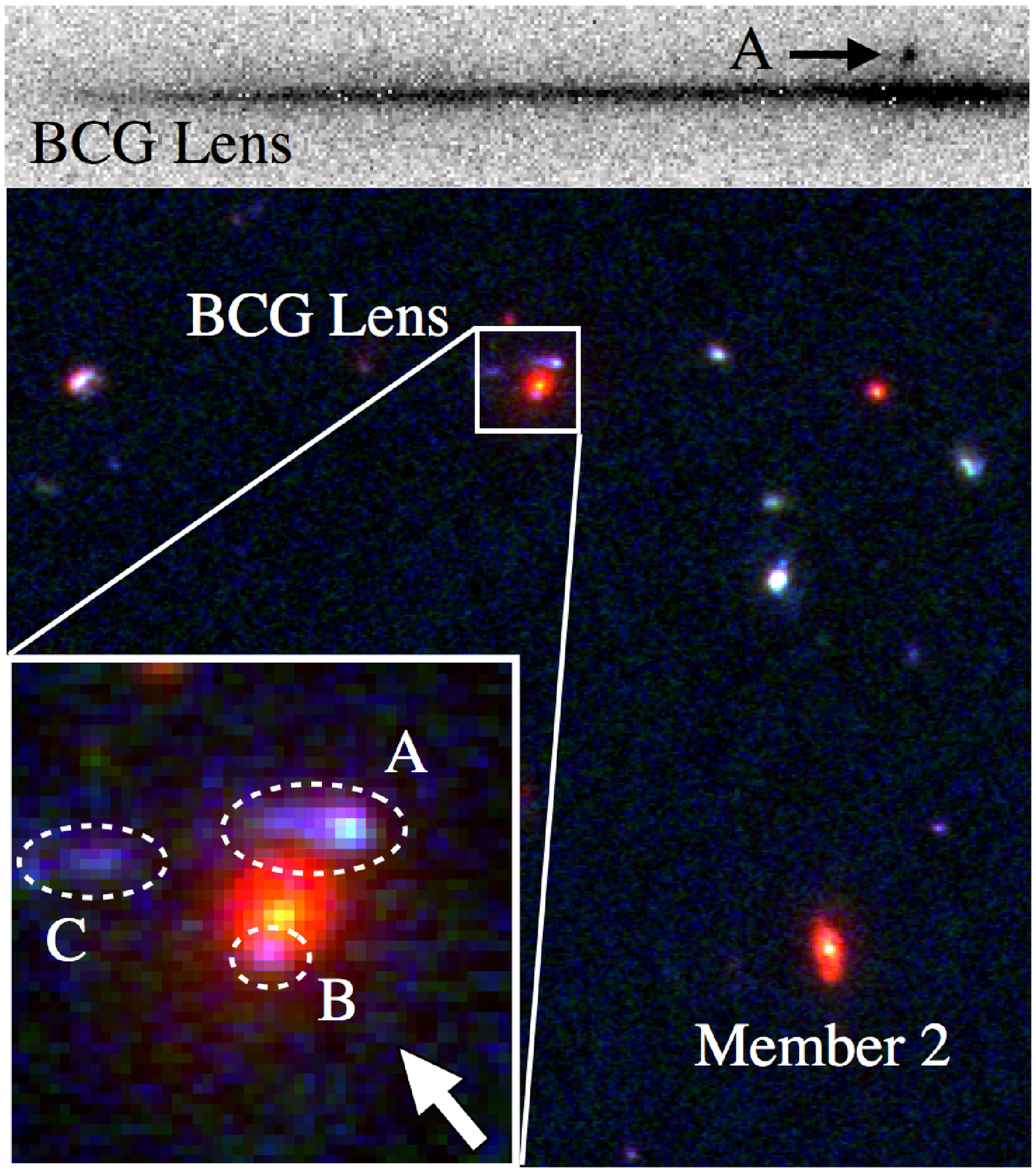}
\caption{{\bf Top:} Negative image of the \hubble\ WFC3/G141 grism
spectrum for the lensing galaxy (continuum).  Approximately $\sim0.5\arcsec$ above the lens is 
\OIII\ emission from object A at \zsource.  {\bf Bottom:} False-color
image of the UDS cluster core generated with \hubble\ imaging
(ACS/F475W, ACS/F814W, WFC3/F125W, and WFC3/F160W).  North and
East are up and left, respectively.  Labeled are the BCG lens and the
other BCG (Member 2).  The inset ($3\arcsec \times 3\arcsec$) shows
the lens with the arc (A) and counterimage (B) labeled.  Another faint galaxy (C) is weakly detected by the grism and consistent with $z\sim2.26$. The arrow indicates the dispersion direction of the grism observations (top).
\label{fig:uds-color}}
\end{figure}

The lens and source are blended in ground-based observations, but
\hubble's resolution (Figure~\ref{fig:uds-color}) separates the system
into the BCG, an arc (object A), and a counterimage (object B).  In the grism spectra, the BCG shows strong continuum.
Approximately $0.5\arcsec$ above the BCG in the G141 spectrum is a compact
emission line corresponding to \OIII\ at \zsourcefull~from object A.  We
also detect \OIII\ corresponding to object B once we remove the lens galaxy's
light (Momcheva et al. in preparation).  The faint object C shows weak
\OIII\ in the grism spectrum consistent with a redshift of \zsource.

\section{Modeling the Strong Lensing System} \label{sec:model}

Combining our ground-based spectroscopy and \hubble~
observations, we confirm that the strong lensing system is composed of the BCG lens and source.  We model the system with {\sc Glee}, a software developed by
S.~H.~Suyu and A.~Halkola \citep{suyu2010, suyu2012}.  Lensing mass
distributions are parametrized profiles, and background
sources are modeled on a pixel grid \citep{suyu2006}.  The lens galaxy light distributions are modeled as S\'ersic profiles.  Unresolved point sources on the pixel grid are
modeled as point images on the image plane.  Model parameters of the
lens and the source are constrained through Markov Chain Monte Carlo
(MCMC) sampling.

\subsection{Lens Model} \label{subsec:lensmod}

\begin{figure*}
\centering
\plotone{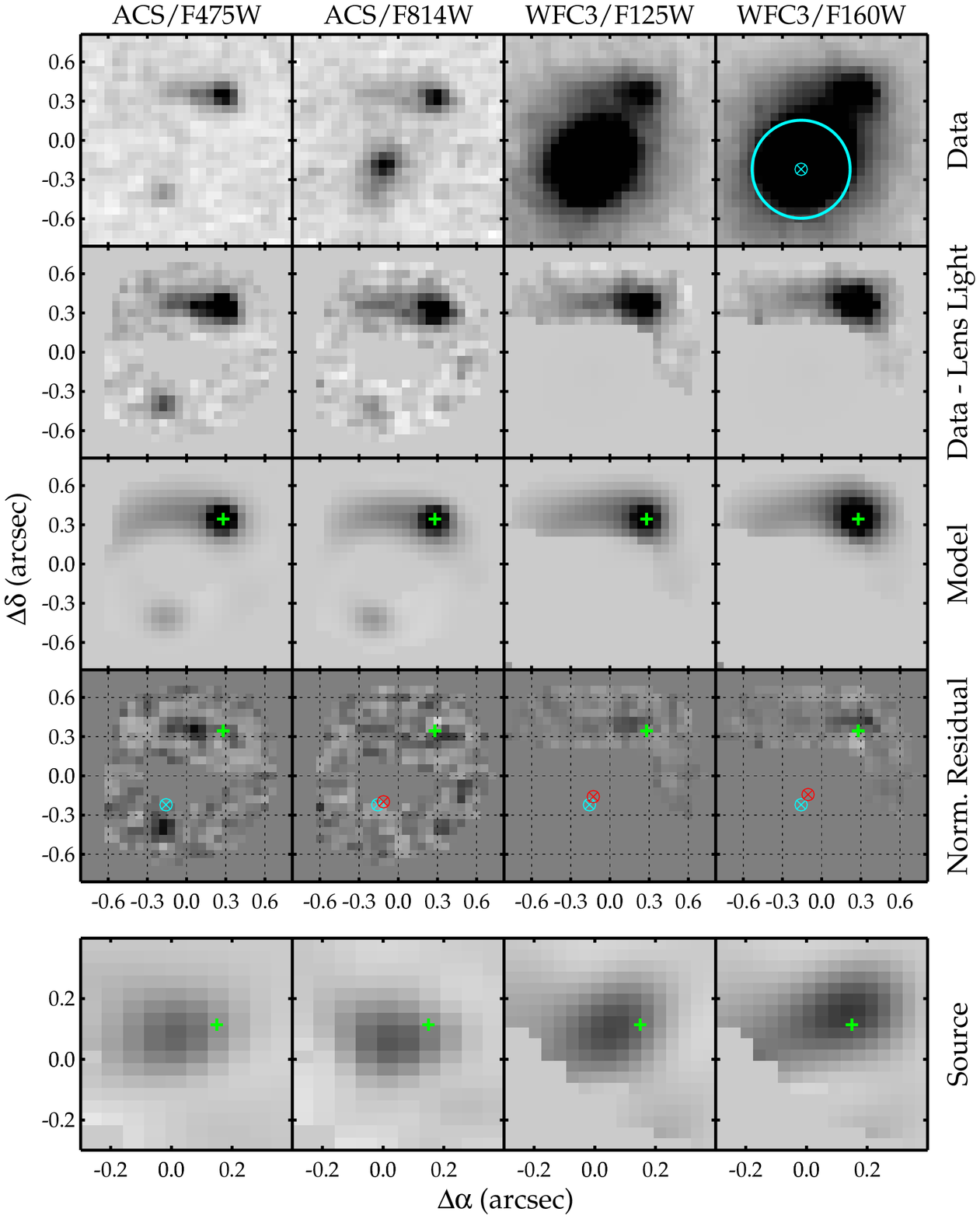}
\caption{{\bf Top Row:} Negative \hubble\ images of the strong
lensing system in the ACS/F475W, ACS/F814W, WFC3/F125W, and WFC3/F160W filters.  The
images are 1.62\arcsec~on a side.  The cyan cross denotes the centroid
of the mass distribution, and the cyan circle the Einstein radius.
{\bf Second Row}: Images with the lens light subtracted.  Only the pixels containing the lensed source are shown and used in the lens modeling.
{\bf Third Row}: Reconstructed image plane from the best-fit lens model.  The model includes a compact emission region (point source convolved with the PSF) marked by the green cross.
{\bf Fourth Row}: Normalized image plane residuals where the color
scale shows the $\pm 2\sigma$ range.  The green cross denotes the
position of the compact source as in the third row.  The magnification varies by $\sim0.5$ over the region around the compact source.  Grid lines are overplotted for clarity.  The centroid of the lens light distribution in each filter (red cross) is offset from that of the mass (cyan cross).
{\bf Bottom Row}: Extended source reconstruction.  The compact emission region is not reconstructed since it is unresolved, but its inferred location from the lens model is indicated (green cross).  The source morphology is less constrained in the F125W and F160W because
only the upper half-annulus is used in the model.  The compact emission is offset from the source galaxy's center in all filters.
\label{fig:lensdata_noshear}}
\end{figure*}

Taking a $27 \times 27$ pixel region around the lens, we
utilize the deepest \hubble\ images spanning the widest wavelength
range: ACS/F475W, ACS/F814W, WFC3/F125W, and WFC3/F160W
(Figure~\ref{fig:lensdata_noshear}, top rows).  We model the lens
galaxy as an elliptical singular power-law mass distribution with the
mass profile slope parameterized as $\Gamma=(\gamma'-1)/2$
(where the 3-dimensional mass density is $\rho(r)\propto
r^{-\gamma'}$).  We adopt a uniform prior of $0.2 \leq
\Gamma \leq 0.8$.  The projected axis ratio $b/a$ has a uniform prior
($0.3 \leq b/a \leq 1$) with the position angle as a free
parameter.  The normalization of the mass profile is set by the
Einstein radius $\theta_{\rm E}$, which is also a free parameter.  We
adopt a Gaussian prior on the mass centroid with a width of $\sigma =
0.05\arcsec$ \citep[see][]{koopmans2006} that is centered on the
fitted light centroid of the lens galaxy in the WFC3/F160W image
($\lambda_{\rm rest}\sim0.6~\mu {\rm m}$).

We first remove the lens galaxy's light using
a single S\'ersic profile, as this is sufficient to describe the lens light in the region of the lensed images.  In F814W, we exclude the annular region containing the lensed images (objects A \& B; Figure~\ref{fig:uds-color}) when fitting the lens light profile.  
Because the lens is considerably brighter than the counterimage (object B;
Figure~\ref{fig:uds-color}) in F125W and F160W, we exclude only the
half-annulus around arc A when fitting the lens light profile.

To constrain the lens mass parameters, we use image pixels
in the annular regions in F475W and F814W and the
half-annular regions in F125W and F160W (Figure~\ref{fig:lensdata_noshear}, second row) that contain the lensed images.  
We construct weight images for each filter by adding Poisson noise from sources to the inverse variance images, as outlined in \citet{koekemoer2011} for CANDELS.  
{\sc Glee} simultaneously models across the four bands and reconstructs the source onto a grid
of $20\times20$ pixels with a resolution of $\sim 0.05''$.  Our lens model also requires a point source convolved with the
point-spread function (PSF) coincident with the peak surface
brightness in image A (Figure~\ref{fig:uds-color}).  The position of this point source is the same across all filters and fit simultaneously with the lens model
parameters.  The counterimage of the point source is not modeled separately given its
low magnification.

\subsection{Modeling Lens Environment Contributions} \label{subsec:env}

The lens galaxy is embedded in a cluster, and the overdense environment may affect the lens model \citep[e.g.,][]{momcheva2006,wong2011}.  We model cluster
galaxies within 2\arcmin~of the lens as singular isothermal
spheres truncated at $r_{200}$ and estimate their virial
masses assuming the stellar-to-halo mass
relation of \citet{moster2013}.  We parameterize the cluster's
dark matter halo as a spherical NFW profile \citep{navarro1996} with a mass of
$7.7\times10^{13}$~\Msun~centered at the peak of the X-ray emission \citep[source 12A
in][]{pierre2012} and a concentration calculated using the mass-concentration relation of \citet{zhao2009}.

In addition to this fiducial model, we run an extensive suite of
models to test environmental contributions to the lens.  We explore multiple permutations with and without: (i) nearby cluster
galaxies; (ii) the cluster dark matter
halo; (iii) additional external shear; (iv) galaxies along the line of
sight, assumed to be in the lens plane.  We also run test models that
allow the cluster halo centroid to vary by the $25\arcsec$ uncertainty
on the X-ray emission centroid \citep{pierre2012}, as well as models where we double the cluster mass.

In exploring this wide variety of models, we find that they provide fits of comparable quality, and the inferred parameters (e.g., $\theta_{E}$) among the different models agree to within their statistical uncertainties.  We adopt our fiducial power-law ellipsoid lens model with a fixed NFW halo for the cluster and fixed isothermal halos for the cluster galaxies because it is realistic and has the fewest additional free parameters.

\section{Results} \label{sec:results}

\subsection{Properties of the Lensing Galaxy} \label{subsec:lens_results}

With our fiducial lens model, we measure an Einstein
radius of $\theta_{\rm E} = 0.38^{+0.02}_{-0.01}\arcsec$ ($3.2_{-0.1}^{+0.2}$ kpc) and total mass enclosed within $\theta_{\rm E}$ of $1.8^{+0.2}_{-0.1}\times
10^{11}$~\Msun.  From our environment model, we estimate that the cluster halo and other cluster members contribute $\sim10\%$ of this enclosed mass ($\sim7\%$ and $\sim3$\%, respectively), consistent with
results at $z\lesssim0.5$ \citep{treu2009}.  Increasing the environment contribution would increase this fraction (e.g., a cluster halo with double the mass would contribute $\sim15\%$ of the enclosed mass), but $\theta_{E}$ remains robust to within the model uncertainties.  The lens galaxy's mass within $\theta_{{\rm E}}$ is thus $\sim1.6\times 10^{11}$~\Msun.

\renewcommand*\arraystretch{1.5}
\begin{table}
\caption{Observed Quantities \& Lens Model Parameters \label{tab:params}}
\begin{ruledtabular}
\begin{tabular}{lr}
\colhead{Parameter} & \colhead{Value} \\
\hline
\multicolumn{2}{c}{\it Observed Quantities}\\
$\alpha, \delta$ [J2000] & 02:18:21.5, -05:10:19.9\\
$z_{\rm L}$ [grism] & $1.6406 ^{+0.0018}_{-0.0050}$\\
$z_{\rm S}$ [grism] & $2.2623 \pm 0.0002$ \\
$z_{\rm S}$ [\lya] & $2.25384 \pm 0.00003$ \\
\lya\ EW$_{\rm observed} (\mathrm{\AA})$ & $129.4 \pm 4.6$ \\
WFC3/F125W (mag)\tablenotemark{a} & $21.19\pm0.01$\\
WFC3/F160W (mag)\tablenotemark{a} & $20.67\pm0.01$\\
M$_{\star}^{{\rm Chab}}$ ($10^{11}$ \Msun)\tablenotemark{b} & $2.0^{+0.8}_{-0.3}$\\
M$_{\star}^{{\rm Salp}}$ ($10^{11}$ \Msun)\tablenotemark{b} & $3.5^{+1.4}_{-0.6}$\\
Bulge+Disk profile [F160W] & \\
~~~Bulge ($n=4$) $r_{\rm eff}$ (\arcsec) & $0.15_{-0.03}^{+0.04}$\\
~~~Disk  ($n=1$) $r_{\rm eff}$ (\arcsec) & $0.72_{-0.07}^{+0.10}$\\
\hline
\multicolumn{2}{c}{\it Lens Model: Galaxy Parameters\tablenotemark{c}} \\
$\mu_{{\rm tot}}$ & $2.1_{-0.3}^{+0.4}$ \\
$|\mu_{A}|$\tablenotemark{d} & $1.8_{-0.2}^{+0.3}$ \\
$|\mu_{B}|$\tablenotemark{d} & $0.3_{-0.1}^{+0.1}$ \\
$\Delta x$ (\arcsec)\tablenotemark{e} & $-0.05_{-0.02}^{+0.01}$ \\
$\Delta y$ (\arcsec)\tablenotemark{e} & $-0.08_{-0.02}^{+0.02}$ \\
$\theta_{\rm E}$ (\arcsec) & $0.38_{-0.01}^{+0.02}$ \\
$b/a$ & $0.8_{-0.2}^{+0.1}$ \\
$\theta~(^{\circ})$  & $8_{-41}^{+45}$ \\
$\Gamma$ & $0.70_{-0.07}^{+0.06}$\\
${\rm M}_{\rm tot}(< \theta_{\rm E})$ ($10^{11}$ \Msun) & $1.8_{-0.1}^{+0.2}$\\
${\rm M}_{\star}^{{\rm Chab}}(< \theta_{\rm E})$ ($10^{11}$ \Msun) & $1.3_{-0.2}^{+0.5}$\\
${\rm M}_{\star}^{{\rm Salp}}(< \theta_{\rm E})$ ($10^{11}$ \Msun) & $2.2_{-0.4}^{+0.9}$\\
$f_{\rm DM}^{{\rm Chab}} (<\theta_{\rm E})$ & $0.3_{-0.3}^{+0.1}$ \\
$f_{\rm DM}^{{\rm Salp}} (<\theta_{\rm E})$ & $-0.3_{-0.5}^{+0.2}$ \\
\hline
\multicolumn{2}{c}{\it Reconstructed Source: Intrinsic Values\tablenotemark{c}} \\
ACS/F475W (mag) & $25.5\pm 0.2$\\
ACS/F814W (mag) & $25.5\pm 0.2$ \\
WFC3/F125W (mag) & $25.9\pm 0.2$ \\
WFC3/F160W (mag) & $25.2\pm 0.2$\\
\end{tabular}
\end{ruledtabular}
\tablenotetext{1}{Total magnitudes from \citet{skelton2014}.}
\tablenotetext{2}{Total stellar masses from \citet{papovich2012}.}
\tablenotetext{3}{For lens model quantities, the reported values are medians, with errors corresponding to the 16th and 84th percentiles.}
\tablenotetext{4}{$\mu_{A}, \mu_{B}$ are magnifications at the peak locations of the brighter and fainter images, respectively.}
\tablenotetext{5}{Offset of mass profile centroid relative to galaxy light centroid in F160W.}
\end{table}
\renewcommand*\arraystretch{1.0}

Following \citet{papovich2012}, we now model the lens light in F160W with a two-component model using an $n=4$ bulge and $n=1$ disk.  The arc is masked, although the source flux is negligible compared to the lens flux ($<3\%$).  \citet{papovich2012} measure the total stellar mass assuming a \citet{chabrier2003} IMF with solar metallicity.  We integrate the light profile and calculate the stellar mass within $\theta_{{\rm E}}$ to be $1.3_{-0.2}^{+0.5} \times 10^{11}~{\rm M}_{\odot}$, corresponding to a dark matter fraction of $f_{\rm DM}^{{\rm Chab}} (< \theta_{\rm E}) = 0.3_{-0.3}^{+0.1}$.  If we use a \citet{salpeter1955} IMF, which seems to be preferred for ETGs at $z<1$ \citep[e.g.,][]{auger2010,cappellari2012,sonnenfeld2012,conroy2013}, the enclosed stellar mass is $2.2_{-0.4}^{+0.9} \times 10^{11}~{\rm M}_{\odot}$ and is larger than the enclosed total mass ($f_{\rm DM}^{{\rm Salp}} (< \theta_{\rm E}) = -0.3_{-0.5}^{+0.2}$), albeit at marginal significance.  This is an upper limit on the dark matter fraction of the galaxy, as the environment contributes $\sim 10\%$ of the total mass within $\theta_{{\rm E}}$.

In the three reddest filters (the lens is undetected in F475W), the centroids of the lens light profile and of the mass model are offset by as much as $\sim1.5$ pixels ($\sim0.1$\arcsec; Figure~\ref{fig:lensdata_noshear}). Studies of $z\lesssim0.5$ ETGs do not find significant offsets between their mass and light \citep{koopmans2006}.  However, the
lens light centroid shifts among different bands, suggesting that the youngest stars are displaced from the older population.  The different stellar populations may not all directly
trace the dark matter.

\subsection{Properties of the Lensed Source}
\label{subsec:source_results}

The source is magnified by a factor of $2.1_{-0.3}^{+0.4}$.  The unlensed source has an extended profile with a bright compact region that is mapped to
the outskirts of the source (Figure~\ref{fig:lensdata_noshear}).  We fit
a S\'ersic profile to the extended source to measure its flux.  To compute the total
flux (Table~\ref{tab:params}), we add the flux from the compact emission, which is determined by taking the flux of the fitted PSF component and dividing by the
magnification at that location ($|\mu_{A}|$ in
Table~\ref{tab:params}).  We estimate the modeling uncertainty to be
$\sim 0.15$ mag by fitting to a subset of sources
reconstructed from samples in the MCMC chain.  Combined with the
photometric uncertainties, the total uncertainty is $\sim0.2$ mag.  The effective radius varies between $r_{\rm eff}\sim0.11''-0.20''$.

There is residual image flux (Figure~\ref{fig:lensdata_noshear}, fourth row) 
that we attribute to compact emission regions in the source.  The residuals shift in both intensity and position among the bands, likely because of different emission lines (Figure~\ref{fig:sed}, bottom): F814W samples only the continuum
and has the smallest residuals, while F475W has \lya\ and F160W has \OIII\ emission.  Indeed, \lya\ emission can be spatially offset from continuum emission in galaxies at $z > 2$ \citep[e.g.,][]{feldmeier2013,momose2014}.  We speculate that offset \OII\ emission may be causing the weaker residual feature in F125W.  Although the G141 spectrum does not show it, \OII\ would fall in a region where the grism response is dropping, making this interpretation uncertain.  These residuals cannot come from the same source position, as lensing is
achromatic.  The lens model uses the same position of the compact emission
region (green cross, Figure~\ref{fig:lensdata_noshear}) across the
multiple bands, but because they are offset,
the model compromises by placing the PSF between the peak fluxes.
This results in residuals in all bands, even F814W where there is only
continuum.

The largest residual is in F475W and is due to \lya\ emission that is
observed in both the arc and counterimage (objects A \& B in
Figure~\ref{fig:uds-color}; see also Figure~\ref{fig:lensdata_noshear},
left).  To test if the bright \lya\ emission is problematic for the
lens modeling, we fit an additional PSF to the counterimage in F475W
and subtract it {\it a priori}.  Using only the upper half-annulus to
model the lens, we find that the image residuals, inferred lens model
parameters, and modeled source fluxes are nearly identical to that of our
fiducial model.

The residuals corresponding to the emission-line regions extend in
the direction of the arc (object A), i.e. tangential to the lens galaxy. 
This implies that the emission-line regions are elongated and/or stretched due to lensing.

\section{Conclusions} \label{sec:conclusions}

We report the discovery of the highest-redshift strong lensing galaxy
known to date.  The unusual gravitational lens is the most massive member of the galaxy cluster \cluster\ at \zcluster.  We measure spectroscopic redshifts for the lens and source of \zbcgfull\ and \zsourcefull, respectively.  Only with \hubble's resolution are we
able to separate the lensing BCG from the source, which is
multiply-imaged into a bright arc and a counterimage.  
The source shows strong \lya\ emission with an observed equivalent width of $129.4 \pm 4.6$~\AA\ as measured with Keck/LRIS.

We adopt a lens model that combines a
power-law ellipsoid for the lens galaxy with an environmental component (fixed
cluster NFW halo + isothermal halos for cluster galaxies).  The
Einstein radius is $\theta_{\rm E}=0.38^{+0.02}_{-0.01}\arcsec$ ($3.2_{-0.1}^{+0.2}$ kpc) and
the total enclosed mass is M$_{\rm tot}(<\theta_{\rm
E})=1.8^{+0.2}_{-0.1}\times10^{11}$~\Msun.  We estimate that the
cluster environment contributes $\sim10$\% of this
total mass.  Assuming a Chabrier IMF for the lens galaxy, the enclosed stellar mass is M$_{\star} (< \theta_{{\rm E}}) = 1.3^{+0.5}_{-0.2} \times 10^{11}~{\rm M}_{\odot}$ and
the upper limit on the galaxy dark matter fraction is $f_{\rm DM}^{{\rm Chab}} (< \theta_{\rm E}) = 0.3_{-0.3}^{+0.1}$.  Alternatively, a Salpeter IMF is marginally inconsistent with the enclosed mass, $f_{\rm DM}^{{\rm Salp}} (< \theta_{\rm E}) = -0.3_{-0.5}^{+0.2}$.

The reconstructed source has bright compact regions offset from the galaxy's center.  These could be emission-line regions corresponding to \lya~and \OIII\ that likely probe different regions in the source.

Strong lensing galaxies at $z > 1$ directly measure the stellar build-up of massive galaxies (e.g., whether the IMF evolves with redshift), and their number tests
the current cosmological paradigm.  Ongoing and upcoming deep imaging
surveys should identify more lensing candidates to directly study the
dark matter and light distributions of structures when the universe
was $\sim30$\% of its current age.

\acknowledgments

We thank Chuck Keeton and Curtis McCully for helpful discussions.  KCW is supported by an EACOA Fellowship awarded by
the East Asia Core Observatories Association, which consists of the
Academia Sinica Institute of Astronomy and Astrophysics, the National
Astronomical Observatory of Japan, the National Astronomical
Observatory of China, and the Korea Astronomy and Space Science
Institute.  This work was supported by a NASA Keck Award, administered
by the NExSci, and data presented were obtained at the W. M. Keck
Observatory from telescope time allocated to NASA through the
partnership with the California Institute of Technology and the
University of California. The Observatory was made possible by the
generous financial support of the W. M. Keck Foundation.  The authors recognize and acknowledge the very significant cultural role
and reverence that the summit of Mauna Kea has always had within the
indigenous Hawaiian community.  This work is based on observations
taken with the NASA/ESA HST by the 3D-HST Treasury Program (GO 12177
and 12328), as well as by program GO 12590, supported by a NASA
through a grant from the Space Telescope Science Institute.  HST and
STScI are operated by the Association of Universities for Research in
Astronomy, Inc., under NASA contract NAS5-26555.

\bibliography{lenspaper}

\end{document}